\documentclass[sigconf]{acmart}
\usepackage{caption}
\usepackage{subcaption}
\usepackage{tikz}
\usepackage{pgfplots} 
\pgfplotsset{compat=1.16} 
\usepackage{balance}
\usepackage{bigstrut,array,multirow,tabularx}
\usepackage{caption}
\usepackage{dsfont}
\usepackage{mathtools}
\DeclarePairedDelimiter\ceil{\lceil}{\rceil}

\AtBeginDocument{%
  \providecommand\BibTeX{{%
    \normalfont B\kern-0.5em{\scshape i\kern-0.25em b}\kern-0.8em\TeX}}}

\setcopyright{acmcopyright}
\copyrightyear{2022}
\acmYear{2022}
\acmDOI{10.1145/3511808.3557605}

\acmConference[CIKM ’22]{the 31st ACM International Conference on Information and Knowledge Management}{October 17--22,
  2022}{Atlanta, GA, USA}
\acmPrice{15.00}
\acmISBN{978-1-4503-9236-5/22/10}

\begin{document}

\title{\textsf{Grad-Align+}: Empowering Gradual Network Alignment Using Attribute Augmentation}


\author{Jin-Duk Park}
\affiliation{%
  \institution{Yonsei University}
  \city{Seoul}
  \country{Republic of Korea}
}
\email{jindeok6@yonsei.ac.kr}

\author{Cong Tran}
\affiliation{%
  \institution{Posts and Telecommunications Institute of Technology}
  \city{Hanoi}
  \country{Vietnam}
}
\email{congtt@ptit.edu.vn}

\author{Won-Yong Shin}
\authornote{Corresponding author}
\affiliation{%
  \institution{Yonsei University}
  \city{Seoul}
  \country{Republic of Korea}
}
\email{wy.shin@yonsei.ac.kr}

\author{Xin Cao}
\affiliation{%
  \institution{The University of New South Wales Australia}
  \city{Sydney}
  \country{Australia}
}
\email{xin.cao@unsw.edu.au}

\renewcommand{\shortauthors}{Jin-Duk Park, Cong Tran, Won-Yong Shin, \& Xin Cao}
\begin{abstract}
Network alignment (NA) is the task of discovering node correspondences across different networks. Although NA methods have achieved remarkable success in a myriad of scenarios, their satisfactory performance is not without prior anchor link information and/or node attributes, which may not always be available. In this paper, we propose \textsf{Grad-Align+}, a novel NA method using {\it node attribute augmentation} that is quite robust to the absence of such additional information. \textsf{Grad-Align+} is built upon a recent state-of-the-art NA method, the so-called Grad-Align, that {\it gradually} discovers only a part of node pairs until all node pairs are found. Specifically, \textsf{Grad-Align+} is composed of the following key components: 1) augmenting node attributes based on nodes' {\it centrality} measures, 2) calculating an embedding similarity matrix extracted from a graph neural network into which the augmented node attributes are fed, and 3) gradually discovering node pairs by calculating similarities between cross-network nodes with respect to the \textit{aligned cross-network neighbor-pair}. Experimental results demonstrate that \textsf{Grad-Align+} exhibits (a) superiority over benchmark NA methods, (b) empirical validation of our theoretical findings, and (c) the effectiveness of our attribute augmentation module.
\end{abstract}



\maketitle

\section{Introduction}
\label{section 1}

\begin{figure}[t]

        \centering
        \begin{subfigure}[c]{0.4\columnwidth}
                \includegraphics[width=0.90\columnwidth]{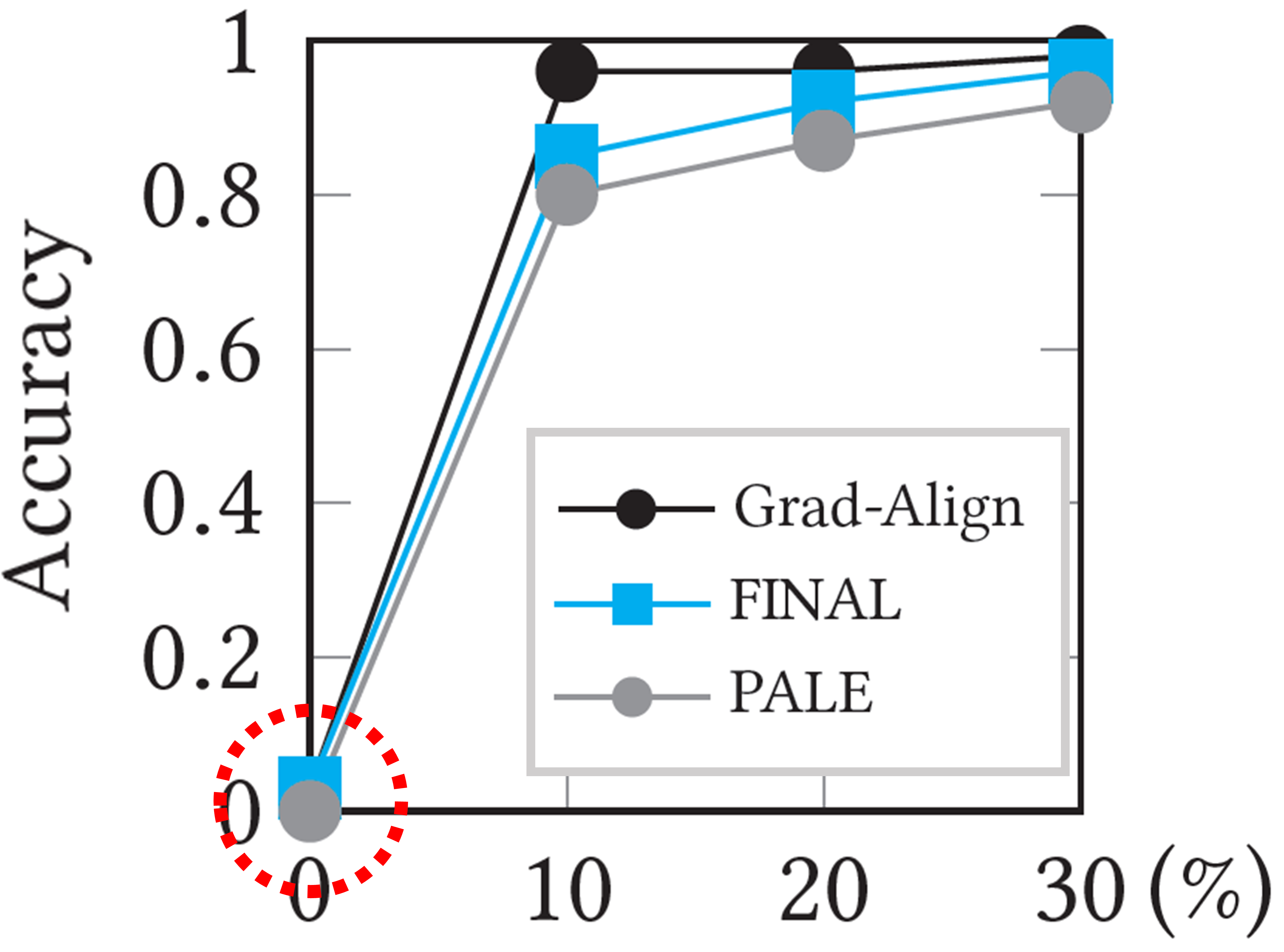}
                \caption{}
                \label{fig1a}
        \end{subfigure}       
        \begin{subfigure}[c]{0.4\columnwidth}
                \includegraphics[width=0.95\columnwidth]{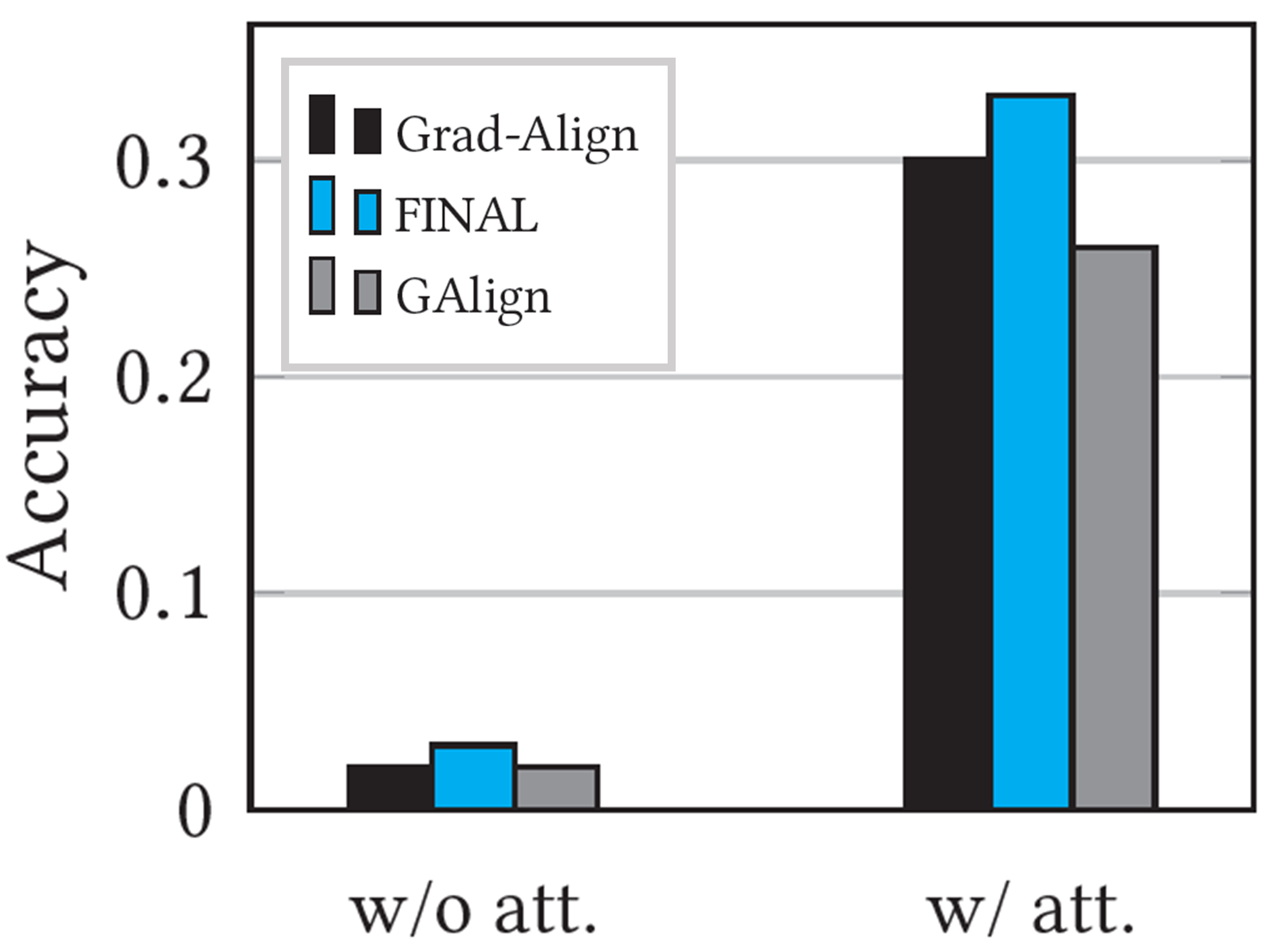}
                \caption{}
                \label{fig1b}
        \end{subfigure}
        \vspace{-1\baselineskip}
        \caption{Alignment accuracy (a) for the scenario where a portion of prior anchor links vary on Facebook vs. Twitter and (b) for the scenario with and without node attributes on Douban Online vs. Douban Offline.}
        \label{fig1} 
\end{figure}

Network alignment (NA) is the task of discovering node correspondences across different networks. NA is often the very first step in performing downstream machine learning (ML) tasks on multiple networks for more precise analyses \cite{zhang2016final, park2022grad, chen2020cone, heimann2018regal}. For example, in social networks, the identification of different accounts (\textit{e.g.}, Facebook, Twitter, and Foursquare) of the same user can facilitate cross-site recommendation, friend recommendation, and personalized advertisement \cite{zhang2016final, trung2020adaptive, zhou2018deeplink}. 

Despite the remarkable success that a number of NA methods \cite{chen2020cone, heimann2018regal, man2016predict, liu2016aligning, zhou2018deeplink, trung2020adaptive, du2019joint, park2022grad,zhang2016final, emmert2016fifty} are achieving in the NA task, their satisfactory performance is not without supervision data ({\it i.e.}, prior anchor link information) and/or node attribute information. Nonetheless, such additional information may not always be available in real-world applications \cite{du2020cross, hsu2017unsupervised, ren2019meta}; in social networks, cross-network anchor link labeling requires tedious user-account pairing and manual user-background checking, which can be very time-consuming and labor-intensive \cite{ren2019meta}.

Our study is motivated by the observation that the state-of-the-art performance of existing NA methods is significantly deteriorated when prior anchor links and node attributes are unavailable. As illustrated in Figure \ref{fig1a}, NA methods designed by leveraging the prior anchor link information ({\it e.g.}, PALE \cite{man2016predict}, FINAL \cite{zhang2016final}, and Grad-Align \cite{trung2020adaptive}) tend to reveal high alignment accuracies when prior anchor links are used (albeit slightly); meanwhile, surprisingly, when anchor link information is no longer available, all of them perform very poorly while showing accuracies much lower than 0.1 (see the red circle depicted in Figure \ref{fig1a}). As illustrated in Figure \ref{fig1b}, NA methods making use of underlying node attributes ({\it e.g.}, GAlign \cite{trung2020adaptive}, FINAL \cite{zhang2016final}, and Grad-Align \cite{park2022grad}) reveal reasonable performance only for attributed networks such as the Douban dataset; thus, their performance drastically degrades for non-attributed network settings with the removal of node attributes.

To tackle this practical challenge, we propose \textsf{Grad-Align+}, a new NA method using {\it node attribute augmentation} that is no longer vulnerable to the absence of additional information such as prior anchor links and node attributes. \textsf{Grad-Align+} is built upon a recent state-of-the-art NA method, named Grad-Align \cite{park2022grad}, that {\it gradually} discovers node pairs by harnessing information enriched through {\it interim} discovery of node correspondences during the node matching. Specifically, in the proposed \textsf{Grad-Align+} method, we first augment node attributes alongside the binning technique based on nodes' {\it centrality} measures. Then, based on the augmented node attributes (and the original node attributes if available), we generate two node representations through two graph neural networks (GNNs) \cite{cui2018survey}, which have emerged as a powerful network feature extractor, trained by using two types of attributes. We then calculate a multi-layer embedding similarity matrix upon the node representations. Finally, we gradually discover node pairs by characterizing a new measure that represents similarities between cross-network nodes with respect to the {\it aligned cross-network neighbor-pair (ACN)} \cite{park2022power} to boost the influence of ACNs during the gradual matching. Through comprehensive experiments using real-world and synthetic datasets, we demonstrate that \textsf{Grad-Align+} (a) enhances the quality of NA, resulting in improving the performance over state-of-the-art NA methods with dramatic gains, (b) empirically validates our theoretical analysis of augmented attributes, and (c) effectively shows the impact and benefits of our attribute augmentation and similarity calculation modules.

\section{Methodology}
\label{section 2}

\subsection{Basic Settings and Problem Definition}
We consider source and target networks to be aligned, denoted as $G_s$ and $G_t$, respectively. For $G_* = (\mathcal{V}_*,\mathcal{E}_*,\mathcal{X}_*)$, the subscript $*$ represents $s$ and $t$ for source and target networks, respectively; $\mathcal{V}_*$ is the set of vertices in $G_*$ whose size is $n_*$; $\mathcal{E}_*$ is the set of edges in $G_*$; and $\mathcal{X}_*$ is the set of original node attributes ({\i.e.}, node metadata) in $G_*$, which is optional.

\begin{definition} (NA)
Given $G_s=(\mathcal{V}_s,\mathcal{E}_s,\mathcal{X}_s)$ and $G_t=(\mathcal{V}_t,\mathcal{E}_t,\mathcal{X}_t)$, NA aims to find one-to-one node mapping $\pi:\mathcal{V}_s \rightarrow \mathcal{V}_t$, where $\pi(u) = v$ and $\pi^{-1}(v) = u$ for $u\in\mathcal{V}_s$ and $v\in\mathcal{V}_t$.
\end{definition}

\begin{figure}[t]
    \centering
    \includegraphics[width=0.36\textwidth]{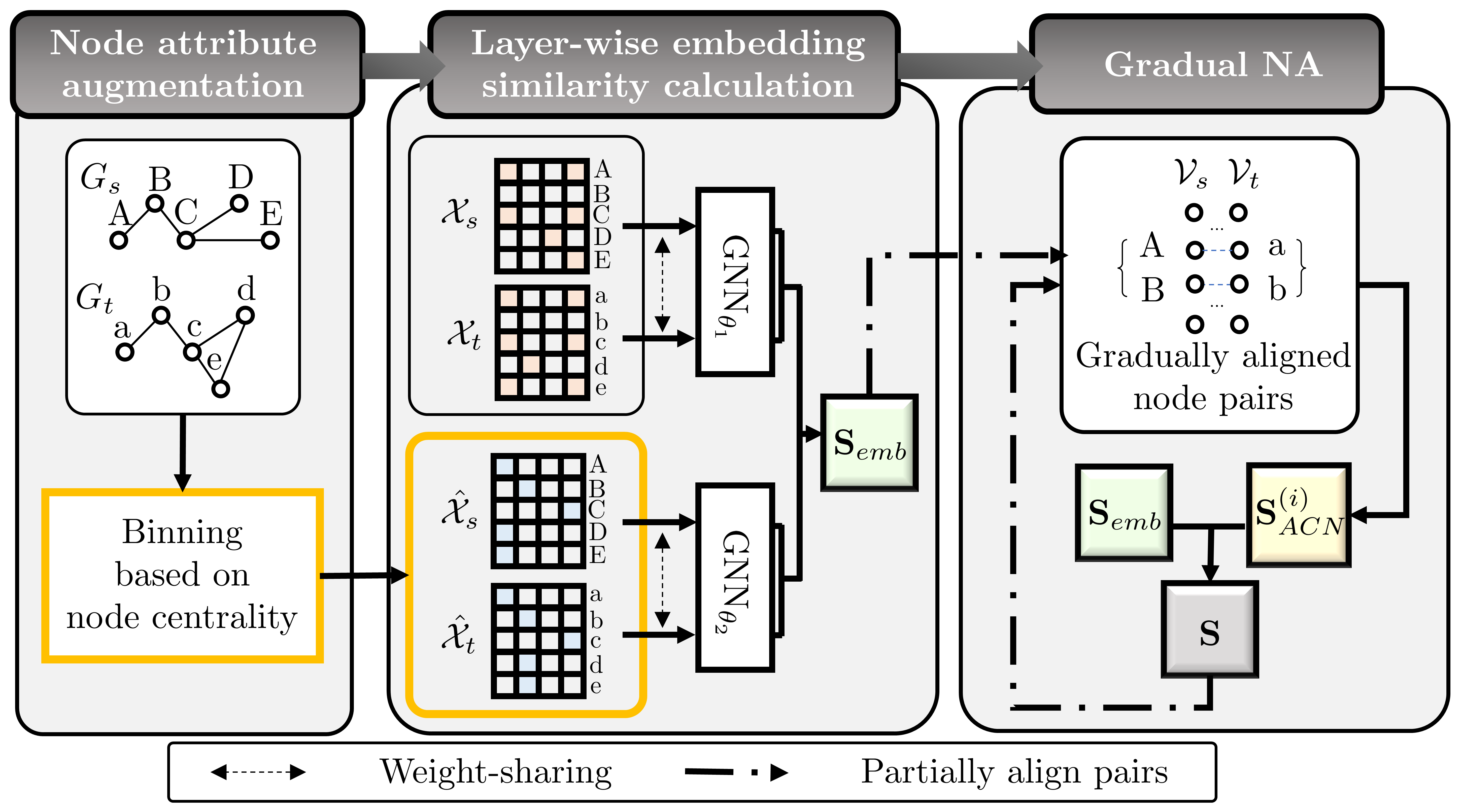}
    \vspace{-3mm}
    \caption{The schematic overview of our \textsf{Grad-Align+} method.}
    \label{fig:overview}
    \setlength{\textfloatsep}{1\baselineskip plus 0.2\baselineskip minus 0.5\baselineskip}
\end{figure}

\subsection{\textsf{Proposed Grad-Align+ Method}}

As illustrated in Figure \ref{fig:overview}, our \textsf{Grad-Align+} is basically composed of three main components: 1) node attribute augmentation, 2) GNN-based embedding similarity calculation, and 3) gradual NA. We elaborate on each component in the following subsections.

\subsubsection{Node Attribute Augmentation}
\label{sec3.step1}
Our node attribute augmentation is built upon the {\it structural consistency} assumption that ground truth cross-network node pairs experience the same degree of influence. This motivates us to augment node attributes based on nodes' {\it centrality} measures. In our study, we adopt $k$-hop centrality \cite{niu2014k} and Katz centrality \cite{katz1953new} among a variety of centrality measures due to their high expressive capability when cross-network node representations are discovered via GNNs using the augmented node attributes.

The $k$-hop centrality is a generalization of degree centrality and measures the influence of all nodes within $k$ hops from the node of interest while the contributions of distant nodes are penalized. The $k$-hop centrality of node $i$ in $G_*$ is formulated as $c_{i,*}=\sum_{l=1}^k \frac{n_{l,*}(i)}{\alpha^{l-1}}$, where $n_{l,*}(i)$ is the number of nodes whose path length from node $i$ is at most $l$ and $\alpha$ is a penalizing constant. On the other hand, the Katz centrality can be viewed as a variant of eigenvector centrality [1] and measures the influence of a node on its higher-order neighbors at larger distances while penalizing higher-order connections. The Katz centrality of node $i$ in $G_*$ is defined as $c_{i,*}=\alpha\sum_{j\in\mathcal{V_*}} a_{ij}c_{j,*}+\beta$, where $a_{ij}$ is the $(i,j)$-th element of the adjacency matrix of $G_*$, and $\alpha$ and $\beta$ are constants. Note that $k$-hop centrality is calculated in a {\it localized} manner while Katz centrality captures both the {\it local and global influences} of a node.

To create node attribute vectors, we discretize the above centrality measures with a fixed dimension $d$. To this end, we employ the equal-width binning technique \cite{catlett1991changing}. Precisely, for node $i$ in $G_*$, we generate the $d$-dimensional one-hot encoded attribute vector, denoted as $\hat{\bf x}_{i,*}$ corresponding to the $i$-th element of matrix $\hat{\mathcal{X}}_*$, in such a way that 1 is assigned to the $\ceil{\frac{c_{i,*}}{w}}$-th element of $\hat{\bf x}_{i,*}$, where $w$ is the binning width. A higher $w$ leads to a lower $d$ due to the fact that $d=\ceil{\frac{{c_{\max,*}}}{w}}$, where ${c_{\max,*}}$ is the maximum centrality among nodes in $G_*$. To avoid a small number of influential nodes to increase the vector dimension, we remove the vector components whose value is never assigned over the two networks. We shall examine how the selection of $d$ affects the performance of \textsf{Grad-Align+} in Section \ref{section3.2}.

\subsubsection{GNN-based Embedding Similarity Calculation}
\label{sec3.step2}
We describe how to calculate the multi-layer embedding similarity matrix via GNNs using augmented node attribute vectors. We start by stating that the augmented node attributes and original node attributes are often {\it heterogeneous} since our augmentation strategy is designed based on the structural information. In this context, instead of na\"{i}vely concatenating two types of attribute vectors, we use two different GNN models, $GNN_{\theta_1}$ and $GNN_{\theta_2}$, into which two attributes are fed separately (see Figure \ref{fig:overview}). The model parameters $\theta_1$ and $\theta_2$ are trained by a \textit{layer-wise reconstruction loss} in \cite{park2022grad} to make each node representation more distinguishable. Using the hidden representations ${\bf H}_*^{(l)}\in\mathbb{R}^{n_*\times h}$ and $\hat{\bf H}_*^{(l)}\in\mathbb{R}^{n_*\times h}$ at each layer extracted from $GNN_{\theta_1}$ and $GNN_{\theta_2}$, respectively, for the dimension $h$ of each vector representation, we are capable of computing the multi-layer embedding similarity matrix as follows:
        \begin{equation}
        \label{emb_sim}
        {\mathbf{S}}_{emb} = \sum_l \mathbf{H}_s^{(l)}{\mathbf{H}_t^{(l)\top}} + \lambda \sum_l  \hat{\mathbf{H}}_s^{(l)}\hat{\mathbf{H}}_t^{(l)\top},
        \end{equation}
where $\lambda$ is a hyperparameter balancing two terms in Eq. (\ref{emb_sim}). For networks without node metadata, we only use the second term.

Next, we analyze how augmented attributed vectors influence the resulting vector representations via GNNs. Let $\hat{\bf h}_{u,s}$ and $\hat{\bf h}_{v,t}$ denote hidden vector representations of nodes $u \in \mathcal{V}_s$ and $v \in \mathcal{V}_t$, respectively. Then, for a ground truth node pair $(u,v)$, it is highly probable to have a small $\|\hat{\bf h}_{u,s}-\hat{\bf h}_{v,t}\|_2$ in the embedding space, where $\|\cdot\|_2$ is the $L_2$-norm of a vector (or a matrix). For ease of analysis, for ground truth node pairs, we make the following two assumptions: 1) $\mathbb{E}[\|\hat{\bf x}_{u,s}-\hat{\bf x}_{v,t}\|_2]$ is arbitrarily small, that is, augmented attributes are consistent, and 2) all the neighbors of $(u,v)$ are ACNs.\footnote{The second assumption is sensical since ground truth node pairs tend to share lots of ACNs in real-world applications \cite{chen2020cone}.} Now, we are ready for establishing the following theorem.

\begin{theorem}
\label{theorem 3.1} 
Consider the pre-activation output of the 1-layer GCN model in which the weight matrix ${\bf W}$ is shared. Suppose that $\mathbb{E}[\|\hat{\bf x}_{u,s}-\hat{\bf x}_{v,t}\|_2] \le \epsilon$ for an arbitrarily small $\epsilon>0$ by the attribute consistency assumption. Then, given a ground truth node pair $(u,v)$ for $u\in\mathcal{V}_s$ and $v\in\mathcal{V}_t$, $\mathbb{E}[\|\hat{\bf h}_{u,s}-\hat{\bf h}_{v,t}\|_2]$ is bounded by $\|{\bf W}\|_2\epsilon$.
\end{theorem}

\subsubsection{Gradual NA}
\label{sec3.step3}
We explain how to gradually match node pairs using a similarity matrix in each gradual step. In this phase, the selection of a similarity measure plays a significant role in determining the performance of NA. In our study, rather than adopting prior approaches based on the Jaccard index \cite{du2019joint} and the Tversky similarity \cite{tversky1977features, park2022power}, as another main contribution, we present our new measure that represents similarities between cross-network nodes, the so-called \textit{ACN similarity}, which is formulated as:
\begin{equation}
\label{ACN_sim}
    {\mathbf S}_{ACN}^{(i)}(u,v) = ACN_{u,v}^p, 
\end{equation}
where ${\bf S}_{ACN}^{(i)}(u,v)$ is the $p$-th power of the number of ACNs between node pair $(u,v)$ for $u\in\mathcal{V}_s$ and $v\in\mathcal{V}_t$ at the $i$-th iteration, corresponding to the $(u,v)$-th element of matrix ${\bf S}_{ACN}^{(i)}$. Finally, following the dual-perception similarity in Grad-Align, we calculate the similarity matrix ${\bf S}^{(i)}$ as follows:
\begin{equation}
\label{overall_sim}
    {\mathbf S}^{(i)} = {\mathbf S}_{emb} \odot {\mathbf S}_{ACN}^{(i)}.
\end{equation}
where $\odot$ indicates the element-wise matrix multiplication operator. The rest of the gradual node matching essentially follows that of \cite{park2022power}.

\section{Experimental evaluation}
\label{section 4}
\subsection{Experimental Setup}
\label{section 4.1}

\noindent\textbf{Datasets.} We conduct experiments on five benchmark datasets consisting of three real-world datasets, including Facebook vs. Twitter (Fb-Tw) \cite{cao2016bass}, Douban Online vs. Douban Offline (Do-Doff) \cite{zhong2012comsoc}, and Allmovie vs. IMDb (Am-ID), and two synthetic datasets, including Facebook and Econ \cite{rossi2015network}. For synthetic datasets, we generate a noisy version of the original network by randomly removing 10\% of edges and flipping 10\% of binary-valued node attributes.

\noindent\textbf{Performance metrics.}
As the most popular metrics, we adopt the {\it alignment accuracy}, denoted as {\it Acc}, and {\it Precision@q} (also known as {\it Success@q}) as in \cite{zhang2016final, du2019joint, park2022power}.

\noindent\textbf{Competitors.} We compare \textsf{Grad-Align+} with 5 state-of-the-art NA methods, which are divided into two different categories: NA methods that necessitate supervision data (\textit{i.e.}, prior anchor links) (FINAL~\cite{zhang2016final} and PALE~\cite{man2016predict}) and NA methods that can be carried out without supervision data (Grad-Align~\cite{park2022grad}, CENALP~\cite{du2019joint}, and GAlign~\cite{trung2020adaptive}).

\noindent\textbf{Implementation details.}
We first describe experimental settings of GNNs. We use GIN \cite{xu2018powerful} as it is validated to exhibit the best performance among well-known GNN models such as GCN \cite{DBLP:conf/iclr/KipfW17} and GraphSAGE \cite{DBLP:conf/nips/HamiltonYL17} (see \cite{park2022power} for more details). We train our GNN model using Adam optimizer \cite{kingma2015adam} with a learning rate of 0.005. For the binning technique, we use 1-hop centrality for the Fb-Tw, Facebook, and Econ datasets and Katz centrality for other datasets unless otherwise stated. For Grad-Align and GAlign, following their original settings, we use all-ones vectors $\mathbf{1} \in \mathbb{R}^{1 \times n_s}$ and $\mathbf{1} \in \mathbb{R}^{1 \times n_t}$ as the input of node attributes. We basically assume \textit{unsupervised settings} where prior anchor links are unavailable. Nevertheless, for the NA methods that should operate on supervision data, we use randomly selected 5\% of prior anchor links as supervision data although our method is handicapped accordingly.

\subsection{Experimental Results}
\label{section3.2}
Our extensive empirical studies are designed to answer the following four key research questions (RQs).
\begin{itemize}
    \item \textbf{RQ1:} How does the choice of centrality in attribute augmentation affect the model performance?
    \item \textbf{RQ2:} How do key parameters affect the performance of \textsf{Grad-Align+}?
    \item \textbf{RQ3:} How much does \textsf{Grad-Align+} improve the NA performance over state-of-the-art NA methods?
    \item \textbf{RQ4:} How much is our attribute augmentation module beneficial to the performance boost of NA methods?
\end{itemize}

\begin{table}[t!]
\footnotesize
\caption{Empirical analysis on node attributes, node representations, and accuracies according to different centrality measures on Facebook.}
\vspace{-1\baselineskip}
\begin{tabular}{cccc}
\hline
Centrality & $\mathbb{E}[\| \mathbf{\hat{x}}_u - \mathbf{\hat{x}}_v \|_2]$ & $\mathbb{E}[\| \mathbf{\hat{h}}_u - \mathbf{\hat{h}}_v \|_2]$ & \textit{Acc} \\ \hline
1-hop     &   \underline{0.53}  &   \underline{5.21}    &   \textbf{84.08}     \\
2-hop     &    0.64 &   5.57    &  50.91        \\
3-hop     &    0.61 &   5.51    &  72.58        \\
Katz     &    \textbf{0.50} &   \textbf{ 5.01}  &  \underline{83.22}       \\ \hline
\end{tabular}
\label{RQ1table}
\vspace{-1\baselineskip}
\end{table}

\noindent \textbf{{(RQ1) Impact of attribute augmentation.}}
We empirically show how node representations behave according to different nodes' centrality measures used for attribute augmentation. To purely examine the impact of augmentation without the original node attributes, we use the Facebook dataset belonging to non-attributed networks in this experiment. From Table \ref{RQ1table}, it is likely that, given ground truth node pairs $(u,v)$ for $u\in\mathcal{V}_s$ and $v\in\mathcal{V}_t$, a lower $\mathbb{E}[\|\hat{\bf x}_{u,s}-\hat{\bf x}_{u,t}\|_2]$ leads to a lower $\mathbb{E}[\|\hat{\bf h}_{u,s}-\hat{\bf h}_{u,t}\|_2]$, thereby resulting in a higher alignment accuracy. This is consistent with Theorem \ref{theorem 3.1} in the sense that $\mathbb{E}[\|\hat{\bf h}_{u,s}-\hat{\bf h}_{u,t}\|_2]$ is bounded by $\mathbb{E}[\|\hat{\bf x}_{u,s}-\hat{\bf x}_{u,t}\|_2]$ to within a constant factor.

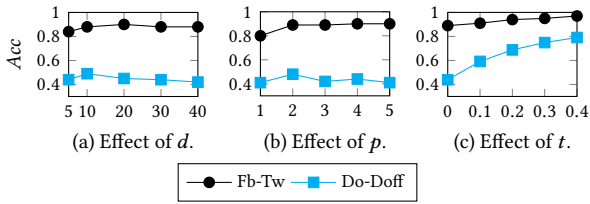
\begin{figure}[t!]
\pgfplotsset{footnotesize,samples=10}
\centering
\begin{tikzpicture}
\begin{axis}[
legend columns=-1,
legend entries={Fb-Tw, Do-Doff},
legend to name=named,
xmax=40,xmin=5,ymin= 0.3,ymax=1,
xlabel=(a) Effect of $d$.,ylabel={\em Acc}, width = 3.3cm, height = 2.7cm,
xtick={5,10,20,30,40},ytick={0.4,0.6,...,1}]
    \addplot+[color=black] coordinates{(5,0.84) (10,0.88) (20,0.90) (30,0.88) (40,0.88) };
    \addplot+[color=cyan] coordinates{(5,0.44) (10,0.49) (20, 0.45) (30,0.44) (40,0.42) };
\end{axis}
\end{tikzpicture}
\begin{tikzpicture}
\begin{axis}[
xmax=5,xmin=1,ymin= 0.3,ymax=1,
xlabel=(b) Effect of $p$., width = 3.3cm, height = 2.7cm,
xtick={1,2,3,4,5},ytick={0.4,0.6,...,1}]
    \addplot+[color=black] coordinates{(1,0.80) (2,0.89) (3,0.89) (4,0.90) (5,0.90) };
    \addplot+[color=cyan] coordinates{(1,0.41) (2,0.48) (3,0.42) (4,0.44) (5,0.41)};
\end{axis}
\end{tikzpicture}
\begin{tikzpicture}
\begin{axis}[
xmax=0.4,xmin=0,ymin= 0.3,ymax=1,
xlabel=(c) Effect of $t$., width = 3.3cm, height = 2.7cm,
xtick={0,0.1,0.2,0.3,0.4},ytick={0.4,0.6,...,1}]
    \addplot+[color=black] coordinates{(0,0.89) (0.1,0.91) (0.2,0.94) (0.3,0.95) (0.4,0.97)};
    \addplot+[color=cyan] coordinates{(0,0.4392) (0.1,0.5921) (0.2,0.6879) (0.3,0.7487) (0.4,0.7907)};
\end{axis}
\end{tikzpicture}
\vspace{-0.5\baselineskip}
\ref{named}
\caption{Alignment accuracy according to different values of parameters used in \textsf{Grad-Align+} and experimental settings.}
\label{Q2plot}
\end{figure}

\noindent\textbf{(RQ2) Effect of key parameters.}
In Figure \ref{Q2plot}, we investigate the effect of parameters used in \textsf{Grad-Align+} and  experimental settings, including the dimension of augmented attribute vectors, $d$, the exponent $p$ of the ACN similarity in Eq. (\ref{ACN_sim}), and the proportion of prior anchor links out of all ground truth pairings, $t$, on the {\it Acc}. We use two real-world datasets exhibiting different degrees of consistency, namely FB-Tw and Do-Doff, which are structurally consistent and structurally inconsistent, respectively \cite{du2019joint}.
 \begin{itemize}
     \item \textbf{The effect of $d$}: Note that the node attributes are augmented based on nodes' centrality ({\it i.e.}, structural information). Thus, it is possible to further enhance the expressiveness of networks having strong structural consistency by increasing the value of $d$. As shown in Figure \ref{Q2plot}a, the performance tends to monotonically increase with $d$ on Fb-Tw while it is rather degraded beyond a certain value of $d$ on Do-Doff due to the excessive use of structural information in networks revealing structural inconsistency.
     \item \textbf{The effect of $p$}: The results from Figure \ref{Q2plot}b essentially show a tendency similar to those in Figure \ref{Q2plot}a. The overexploitation of ACNs during the node matching on Do-Doff exhibiting strong structural inconsistency turns out to degrade the performance.
     \item \textbf{The effect of $t$}: From Figure \ref{Q2plot}c, it is obvious that the {\it Acc} monotonically increases with more supervision data. One can see that \textsf{Grad-Align+} still guarantees quite reasonable performance even in unsupervised settings ({\it i.e.}, $t=0$) unlike existing NA methods (refer to Figure \ref{fig1}).
 \end{itemize}


\begin{table}[t!]
\caption{Performance comparison among \textsf{Grad-Align+} and 5 state-of-the-art NA methods in terms of the {\it Acc} and {\it Precision@q}. Here, the best and second best performers are highlighted by bold and underline, respectively.}
\centering
\setlength\tabcolsep{3pt} 
\scalebox{0.7}{
\begin{tabular}{ccccccccc}
\toprule 
Method & Metric &Fb-Tw & Do-Doff& Am-ID & Facebook & Econ & \textit{Unsup.} \\
\bottomrule
\multirow{3}{*}{\rotatebox{0}{PALE}} 
  & {\em Acc} & 0.5923  & 0.1052 & 0.5323 & \underline{0.6079} & 0.5819 & \\
  & $Precision@5$ & 0.6851  & 0.1783 & 0.6244 & 0.6040 & 0.6041 &  \\
  & $Precision@10$ & 0.7303  & 0.2338 & 0.7519 & 0.7162 & 0.6836 & \\
\midrule
\multirow{3}{*}{\rotatebox{0}{FINAL}} 
  & {\em Acc} & 0.6328  & 0.2773 & 0.6125 & 0.5780 & 0.4185 &  \\
  & $Precision@5$ & 0.6475  & 0.4358  &0.7592 & 0.5509 & 0.4467 &  \\
  & $Precision@10$ & 0.7253  & 0.5824  & 0.8152 & \underline{0.7199} & 0.5841&  \\
\midrule
\multirow{3}{*}{\rotatebox{0}{CENALP}} 
  & {\em Acc} & \underline{0.9105} &0.0235 & 0.4238 &0.4237 & 0.4872  &  \\
  & $Precision@5$ & \underline{0.9352}  &  0.0571 & 0.5721 & \underline{0.6278} & 0.5328 & \checkmark \\
  & $Precision@10$ & \underline{0.9405} & 0.1130 & 0.7154 & 0.7115 & 0.6271 &  \\
\midrule
\multirow{3}{*}{\rotatebox{0}{GAlign}} 
  & {\em Acc} & 0.0513  & 0.2568 & 0.7364 & 0.0413 & 0.8108 &  \\
  & $Precision@5$ & 0.0422 & 0.5233 & 0.8101 & 0.0622 & 0.8617 & \checkmark \\
  & $Precision@10$ & 0.0612  & 0.6324 & 0.8749 & 0.0899 & 0.8995 & \\
\midrule
\multirow{3}{*}{\rotatebox{0}{Grad-Align}} 
  & {\em Acc} &  0.0218 & \underline{0.2987} & \underline{0.8316} & 0.0231 & \underline{0.8167} &  \\
  & $Precision@5$ & 0.0325  &\underline{0.5707} & \underline{0.9101} & 0.0378 & \underline{0.9097} & \checkmark \\
  & $Precision@10$ & 0.0421 &  \underline{0.6494}  & \underline{0.9308} & 0.0405 & \underline{0.9336} & \\
\midrule
\multirow{3}{*}{\rotatebox{0}{\textsf{Grad-Align+}}} 
  & {\em Acc} & \textbf{0.9156}  &  \textbf{0.4392} & \textbf{0.9318}  & \textbf{0.8408} & \textbf{0.8418} &  \\
  & $Precision@5$ & \textbf{0.9386}  &\textbf{0.6422} &\textbf{0.9640} & \textbf{0.8782} &\textbf{0.9563} & \checkmark\\
  & $Precision@10$ & \textbf{0.9482} & \textbf{0.7308} &\textbf{0.9879} &\textbf{0.8932} & \textbf{0.9690} &  \\
\bottomrule
\end{tabular}}
\vspace{-1mm}
\label{Q3table}
\end{table}

\begin{table}[t!]
\caption{Performance comparison of NA methods with and without our attribute augmentation in terms of the {\it Acc} on Facebook. Here, the best and second best performers are highlighted by bold and underline, respectively.}
\footnotesize 
\centering
\vspace{-1\baselineskip}
\begin{tabular}{cccc}
\toprule
Method      & w/o att. aug. ($X$)  & w/ att. aug. ($Y$) & Gain ($\frac{Y-X}{X} \times 100 (\%)$)\\
\midrule
FINAL       & \textbf{0.5780}                & 0.7032     &   21.66 \\
GAlign      & \underline{0.0513}               &   0.7862 &   1432.55 \\
Grad-Align  &  0.0218      &       \underline{0.8105}    &   \textbf{3617.88} \\
\textsf{Grad-Align+} & 0.0253       &         \textbf{0.8408}      &   \underline{3223.32} \\
\bottomrule\end{tabular}
\label{Q4table}
\end{table}

\noindent\textbf{(RQ3) Comparison with five competitors.}
Table \ref{Q3table} presents the performance comparison between \textsf{Grad-Align+} and 5 state-of-the-art NA methods with respect to the {\it Acc} and {\it Precision@q} for $q \in \{5,10\}$ using three real-world and two synthetic datasets. Here, \textit{Unsup.} represents the methods that are run \textit{unsupervisedly} without any prior anchor links. \textsf{Grad-Align+} consistently and significantly outperforms all the competitors regardless of the datasets and the performance metrics while showing gains up to 47.03\% compared to the second-best performer. However, the second-best performer depends on the datasets, which implies that one does not dominate other competitors. More interestingly, existing GNN-based NA methods ({\it i.e.}, Grad-Align and GAlign) perform poorly on the datasets without node attributes such as Fb-Tw and Facebook. This indicates that the augmented node attributes play a crucial role in generating precise representations of each node, thus resulting in a substantial performance improvement.

\noindent\textbf{(RQ4) Impact of our attribute augmentation module.}
Our node attribute augmentation module can also be integrated into other NA methods that leverage attribute information. To investigate the impact of attribute augmentation in such NA methods, we conduct an ablation study by removing this module and summarize the evaluation results in terms of the {\it Acc} using the Facebook dataset in Table \ref{Q4table}. Dramatic gains over the cases without attribute augmentation are achieved when Grad-Align and \textsf{Grad-Align+} are employed. Since the accuracy of {\it interim} discovery of node correspondences is critical in gradually discovering node pairs, attribute augmentation can most benefit both Grad-Align and \textsf{Grad-Align+} by helping discover correct node pairs, especially in the early stage of gradual node matching.

\section{Concluding remarks}
\label{section 5}
In this paper, we aimed to devise a methodology that substantially improves the performance of NA in unsupervised settings. Toward this goal, we proposed \textsf{Grad-Align+}, the high-quality
 NA method that judiciously integrates the GNN model trained along with augmented node attributes based on $k$-hop and Katz centrality measures into the gradual node matching framework. Through extensive experiments, we demonstrated the effectiveness of our augmentation module with its theoretical validity as well as the superiority of \textsf{Grad-Align+} over the state-of-the-art NA method with gains of up to 47.03\%.

\begin{acks}
The work of W.-Y. Shin was supported by the National Research Foundation of Korea (NRF) grant funded by the Korea government (MSIT) (No. 2021R1A2C3004345). The work of X. Cao was supported by ARC DE190100663.
\end{acks}






%


\bibliographystyle{ACM-Reference-Format}
\balance
\bibliography{sample-base, Cameraready}


\begin{thebibliography}{26}


\ifx \showCODEN    \undefined \def \showCODEN     #1{\unskip}     \fi
\ifx \showDOI      \undefined \def \showDOI       #1{#1}\fi
\ifx \showISBNx    \undefined \def \showISBNx     #1{\unskip}     \fi
\ifx \showISBNxiii \undefined \def \showISBNxiii  #1{\unskip}     \fi
\ifx \showISSN     \undefined \def \showISSN      #1{\unskip}     \fi
\ifx \showLCCN     \undefined \def \showLCCN      #1{\unskip}     \fi
\ifx \shownote     \undefined \def \shownote      #1{#1}          \fi
\ifx \showarticletitle \undefined \def \showarticletitle #1{#1}   \fi
\ifx \showURL      \undefined \def \showURL       {\relax}        \fi
\providecommand\bibfield[2]{#2}
\providecommand\bibinfo[2]{#2}
\providecommand\natexlab[1]{#1}
\providecommand\showeprint[2][]{arXiv:#2}

\bibitem[Cao and Yu(2016)]%
        {cao2016bass}
\bibfield{author}{\bibinfo{person}{Xuezhi Cao} {and} \bibinfo{person}{Yong
  Yu}.} \bibinfo{year}{2016}\natexlab{}.
\newblock \showarticletitle{{BASS}: A bootstrapping approach for aligning
  heterogenous social networks}. In \bibinfo{booktitle}{\emph{Proc. Joint Eur.
  Conf. Mach. Learn. Knowl. Discovery Databases (ECML-PKDD'16)}}.
  \bibinfo{address}{Riva del Garda, Italy}, \bibinfo{pages}{459--475}.
\newblock


\bibitem[Catlett(1991)]%
        {catlett1991changing}
\bibfield{author}{\bibinfo{person}{Jason Catlett}.}
  \bibinfo{year}{1991}\natexlab{}.
\newblock \showarticletitle{On changing continuous attributes into ordered
  discrete attributes}. In \bibinfo{booktitle}{\emph{Proc. Eur. Working Session
  on Learning. (EWSL)}}. \bibinfo{address}{Porto Portugal},
  \bibinfo{pages}{164--178}.
\newblock


\bibitem[Chen et~al\mbox{.}(2020)]%
        {chen2020cone}
\bibfield{author}{\bibinfo{person}{Xiyuan Chen}, \bibinfo{person}{Mark
  Heimann}, \bibinfo{person}{Fatemeh Vahedian}, {and} \bibinfo{person}{Danai
  Koutra}.} \bibinfo{year}{2020}\natexlab{}.
\newblock \showarticletitle{{CONE-Align}: Consistent network alignment with
  proximity-preserving node embedding}. In \bibinfo{booktitle}{\emph{Proc. 29th
  ACM Int. Conf. Inf. Knowl. Manage. (CIKM'20)}}. \bibinfo{address}{Virtual
  Event}, \bibinfo{pages}{1985--1988}.
\newblock


\bibitem[Cui et~al\mbox{.}(2019)]%
        {cui2018survey}
\bibfield{author}{\bibinfo{person}{Peng Cui}, \bibinfo{person}{Xiao Wang},
  \bibinfo{person}{Jian Pei}, {and} \bibinfo{person}{Wenwu Zhu}.}
  \bibinfo{year}{2019}\natexlab{}.
\newblock \showarticletitle{A survey on network embedding}.
\newblock \bibinfo{journal}{\emph{IEEE Trans. Knowl. Data Eng.}}
  \bibinfo{volume}{31}, \bibinfo{number}{5} (\bibinfo{date}{May}
  \bibinfo{year}{2019}), \bibinfo{pages}{833--852}.
\newblock


\bibitem[Du et~al\mbox{.}(2019)]%
        {du2019joint}
\bibfield{author}{\bibinfo{person}{Xingbo Du}, \bibinfo{person}{Junchi Yan},
  {and} \bibinfo{person}{Hongyuan Zha}.} \bibinfo{year}{2019}\natexlab{}.
\newblock \showarticletitle{Joint link prediction and network alignment via
  cross-graph embedding}. In \bibinfo{booktitle}{\emph{Proc. 28th Int. Joint
  Conf. Artif. Intell. (IJCAI'19)}}. \bibinfo{address}{Macao, China},
  \bibinfo{pages}{2251--2257}.
\newblock


\bibitem[Du et~al\mbox{.}(2020)]%
        {du2020cross}
\bibfield{author}{\bibinfo{person}{Xingbo Du}, \bibinfo{person}{Junchi Yan},
  \bibinfo{person}{Rui Zhang}, {and} \bibinfo{person}{Hongyuan Zha}.}
  \bibinfo{year}{2020}\natexlab{}.
\newblock \showarticletitle{Cross-network skip-gram embedding for joint network
  alignment and link prediction}.
\newblock \bibinfo{journal}{\emph{IEEE Trans. Knowl. Data Eng.}}
  \bibinfo{volume}{34}, \bibinfo{number}{3} (\bibinfo{year}{2020}),
  \bibinfo{pages}{1080--1095}.
\newblock


\bibitem[Emmert-Streib et~al\mbox{.}(2016)]%
        {emmert2016fifty}
\bibfield{author}{\bibinfo{person}{Frank Emmert-Streib},
  \bibinfo{person}{Matthias Dehmer}, {and} \bibinfo{person}{Yongtang Shi}.}
  \bibinfo{year}{2016}\natexlab{}.
\newblock \showarticletitle{Fifty years of graph matching, network alignment
  and network comparison}.
\newblock \bibinfo{journal}{\emph{Inf. Sci.}}  \bibinfo{volume}{346}
  (\bibinfo{date}{Jun.} \bibinfo{year}{2016}), \bibinfo{pages}{180--197}.
\newblock


\bibitem[Hamilton et~al\mbox{.}(2017)]%
        {DBLP:conf/nips/HamiltonYL17}
\bibfield{author}{\bibinfo{person}{Will Hamilton}, \bibinfo{person}{Zhitao
  Ying}, {and} \bibinfo{person}{Jure Leskovec}.}
  \bibinfo{year}{2017}\natexlab{}.
\newblock \showarticletitle{Inductive representation learning on large graphs}.
  In \bibinfo{booktitle}{\emph{Proc. 28th Int. Conf. Neural Inf. Process. Syst.
  (NIPS'17)}}. \bibinfo{address}{Long Beach, CA}, \bibinfo{pages}{1024--1034}.
\newblock


\bibitem[Heimann et~al\mbox{.}(2018)]%
        {heimann2018regal}
\bibfield{author}{\bibinfo{person}{Mark Heimann}, \bibinfo{person}{Haoming
  Shen}, \bibinfo{person}{Tara Safavi}, {and} \bibinfo{person}{Danai Koutra}.}
  \bibinfo{year}{2018}\natexlab{}.
\newblock \showarticletitle{{REGAL}: Representation learning-based graph
  alignment}. In \bibinfo{booktitle}{\emph{Proc. 27th ACM Int. Conf. Inf.
  Knowl. Manage. (CIKM'18)}}. \bibinfo{address}{Turin, Italy},
  \bibinfo{pages}{117--126}.
\newblock


\bibitem[Hsu et~al\mbox{.}(2017)]%
        {hsu2017unsupervised}
\bibfield{author}{\bibinfo{person}{Chin-Chi Hsu}, \bibinfo{person}{Yi-An Lai},
  \bibinfo{person}{Wen-Hao Chen}, \bibinfo{person}{Ming-Han Feng}, {and}
  \bibinfo{person}{Shou-De Lin}.} \bibinfo{year}{2017}\natexlab{}.
\newblock \showarticletitle{Unsupervised ranking using graph structures and
  node attributes}. In \bibinfo{booktitle}{\emph{Proc. 10th ACM Int. Conf. Web
  Search Data Min. (WSDM)}}. \bibinfo{address}{New York, NY},
  \bibinfo{pages}{771–779}.
\newblock


\bibitem[Katz(1953)]%
        {katz1953new}
\bibfield{author}{\bibinfo{person}{Leo Katz}.} \bibinfo{year}{1953}\natexlab{}.
\newblock \showarticletitle{A new status index derived from sociometric
  analysis}.
\newblock \bibinfo{journal}{\emph{Psychometrika}} \bibinfo{volume}{18},
  \bibinfo{number}{1} (\bibinfo{date}{Mar.} \bibinfo{year}{1953}),
  \bibinfo{pages}{39--43}.
\newblock


\bibitem[Kingma and Ba(2015)]%
        {kingma2015adam}
\bibfield{author}{\bibinfo{person}{Diederik~P. Kingma} {and}
  \bibinfo{person}{Jimmy Ba}.} \bibinfo{year}{2015}\natexlab{}.
\newblock \showarticletitle{Adam: {A} method for stochastic optimization}. In
  \bibinfo{booktitle}{\emph{Proc. 3rd Int. Conf. Learn. Representations
  ({ICLR}'15)}}. \bibinfo{address}{San Diego, CA}.
\newblock


\bibitem[Kipf and Welling(2017)]%
        {DBLP:conf/iclr/KipfW17}
\bibfield{author}{\bibinfo{person}{Thomas~N. Kipf} {and} \bibinfo{person}{Max
  Welling}.} \bibinfo{year}{2017}\natexlab{}.
\newblock \showarticletitle{Semi-supervised classification with graph
  convolutional networks}. In \bibinfo{booktitle}{\emph{Proc. 5th Int. Conf.
  Learning Rep. (ICLR'17)}}. \bibinfo{address}{Toulon, France},
  \bibinfo{pages}{1--14}.
\newblock


\bibitem[Liu et~al\mbox{.}(2016)]%
        {liu2016aligning}
\bibfield{author}{\bibinfo{person}{Li Liu}, \bibinfo{person}{William~K Cheung},
  \bibinfo{person}{Xin Li}, {and} \bibinfo{person}{Lejian Liao}.}
  \bibinfo{year}{2016}\natexlab{}.
\newblock \showarticletitle{Aligning users across social networks using network
  embedding.}. In \bibinfo{booktitle}{\emph{Proc. 25th Int. Joint Conf. Artif.
  Intell. (IJCAI'16)}}. \bibinfo{address}{New York City, NY},
  \bibinfo{pages}{1774--1780}.
\newblock


\bibitem[Man et~al\mbox{.}(2016)]%
        {man2016predict}
\bibfield{author}{\bibinfo{person}{Tong Man}, \bibinfo{person}{Huawei Shen},
  \bibinfo{person}{Shenghua Liu}, \bibinfo{person}{Xiaolong Jin}, {and}
  \bibinfo{person}{Xueqi Cheng}.} \bibinfo{year}{2016}\natexlab{}.
\newblock \showarticletitle{Predict anchor links across social networks via an
  embedding approach.}. In \bibinfo{booktitle}{\emph{Proc. 25th Int. Joint
  Conf. Artif. Intell. (IJCAI'16)}}. \bibinfo{address}{New York City, NY},
  \bibinfo{pages}{1823--1829}.
\newblock


\bibitem[Niu et~al\mbox{.}(2014)]%
        {niu2014k}
\bibfield{author}{\bibinfo{person}{Jianwei Niu}, \bibinfo{person}{JinYang Fan},
  \bibinfo{person}{Lei Wang}, {and} \bibinfo{person}{Milica Stojinenovic}.}
  \bibinfo{year}{2014}\natexlab{}.
\newblock \showarticletitle{$k$-hop centrality metric for identifying
  influential spreaders in dynamic large-scale social networks}. In
  \bibinfo{booktitle}{\emph{Proc. {IEEE} Global Commun. Conf. (GLOBECOM'14)}}.
  \bibinfo{address}{Austin, Texas}, \bibinfo{pages}{2954--2959}.
\newblock


\bibitem[Park et~al\mbox{.}(2022a)]%
        {park2022grad}
\bibfield{author}{\bibinfo{person}{Jin-Duk Park}, \bibinfo{person}{Cong Tran},
  \bibinfo{person}{Won-Yong Shin}, {and} \bibinfo{person}{Xin Cao}.}
  \bibinfo{year}{2022}\natexlab{a}.
\newblock \showarticletitle{{G}rad-{A}lign: {G}radual network alignment via
  graph neural networks ({S}tudent {A}bstract)}. In
  \bibinfo{booktitle}{\emph{Proc. 36th {AAAI} Conf. on Artif. Intell.
  (AAAI'22)}}. \bibinfo{address}{Virtual Event}.
\newblock


\bibitem[Park et~al\mbox{.}(2022b)]%
        {park2022power}
\bibfield{author}{\bibinfo{person}{Jin-Duk Park}, \bibinfo{person}{Cong Tran},
  \bibinfo{person}{Won-Yong Shin}, {and} \bibinfo{person}{Xin Cao}.}
  \bibinfo{year}{2022}\natexlab{b}.
\newblock \showarticletitle{On the power of gradual network alignment using
  dual-perception similarities}.
\newblock \bibinfo{journal}{\emph{arXiv preprint arXiv:2201.10945}}
  (\bibinfo{year}{2022}).
\newblock


\bibitem[Ren et~al\mbox{.}(2019)]%
        {ren2019meta}
\bibfield{author}{\bibinfo{person}{Yuxiang Ren}, \bibinfo{person}{Charu~C
  Aggarwal}, {and} \bibinfo{person}{Jiawei Zhang}.}
  \bibinfo{year}{2019}\natexlab{}.
\newblock \showarticletitle{Meta diagram based active social networks
  alignment}. In \bibinfo{booktitle}{\emph{Proc. 35th Int. Conf. Data {E}ng.
  (ICDE'19)}}. \bibinfo{address}{Macao, China}, \bibinfo{pages}{1690--1693}.
\newblock


\bibitem[Rossi and Ahmed(2015)]%
        {rossi2015network}
\bibfield{author}{\bibinfo{person}{Ryan Rossi} {and} \bibinfo{person}{Nesreen
  Ahmed}.} \bibinfo{year}{2015}\natexlab{}.
\newblock \showarticletitle{The network data repository with interactive graph
  analytics and visualization}. In \bibinfo{booktitle}{\emph{Proc. 29th {AAAI}
  Conf. on Artif. Intell. (AAAI'15)}}. \bibinfo{address}{Austin, Texas},
  \bibinfo{pages}{4292--4293}.
\newblock


\bibitem[Trung et~al\mbox{.}(2020)]%
        {trung2020adaptive}
\bibfield{author}{\bibinfo{person}{Huynh~Thanh Trung}, \bibinfo{person}{Tong
  Van~Vinh}, \bibinfo{person}{Nguyen~Thanh Tam}, \bibinfo{person}{Hongzhi Yin},
  \bibinfo{person}{Matthias Weidlich}, {and} \bibinfo{person}{Nguyen Quoc~Viet
  Hung}.} \bibinfo{year}{2020}\natexlab{}.
\newblock \showarticletitle{Adaptive network alignment with unsupervised and
  multi-order convolutional networks}. In \bibinfo{booktitle}{\emph{Proc. 36th
  Int. Conf. Data {E}ng. (ICDE'20)}}. \bibinfo{address}{Dallas, TX},
  \bibinfo{pages}{85--96}.
\newblock


\bibitem[Tversky(1977)]%
        {tversky1977features}
\bibfield{author}{\bibinfo{person}{Amos Tversky}.}
  \bibinfo{year}{1977}\natexlab{}.
\newblock \showarticletitle{Features of similarity.}
\newblock \bibinfo{journal}{\emph{Psychol. Rev.}} \bibinfo{volume}{84},
  \bibinfo{number}{4} (\bibinfo{year}{1977}), \bibinfo{pages}{327--352}.
\newblock


\bibitem[Xu et~al\mbox{.}(2019)]%
        {xu2018powerful}
\bibfield{author}{\bibinfo{person}{Keyulu Xu}, \bibinfo{person}{Weihua Hu},
  \bibinfo{person}{Jure Leskovec}, {and} \bibinfo{person}{Stefanie Jegelka}.}
  \bibinfo{year}{2019}\natexlab{}.
\newblock \showarticletitle{How powerful are graph neural networks?}. In
  \bibinfo{booktitle}{\emph{Proc. 7th Int. Conf. Learning Rep. (ICLR'19)}}.
  \bibinfo{address}{New Orleans, LA}.
\newblock


\bibitem[Zhang and Tong(2016)]%
        {zhang2016final}
\bibfield{author}{\bibinfo{person}{Si Zhang} {and} \bibinfo{person}{Hanghang
  Tong}.} \bibinfo{year}{2016}\natexlab{}.
\newblock \showarticletitle{{FINAL}: Fast attributed network alignment}. In
  \bibinfo{booktitle}{\emph{Proc. 22nd ACM SIGKDD Int. Conf. Knowledge
  {D}iscovery {\&} {D}ata {M}ining (KDD'16)}}. \bibinfo{address}{San Francisco,
  CA}, \bibinfo{pages}{1345--1354}.
\newblock


\bibitem[Zhong et~al\mbox{.}(2012)]%
        {zhong2012comsoc}
\bibfield{author}{\bibinfo{person}{Erheng Zhong}, \bibinfo{person}{Wei Fan},
  \bibinfo{person}{Junwei Wang}, \bibinfo{person}{Lei Xiao}, {and}
  \bibinfo{person}{Yong Li}.} \bibinfo{year}{2012}\natexlab{}.
\newblock \showarticletitle{Com{S}oc: adaptive transfer of user behaviors over
  composite social network}. In \bibinfo{booktitle}{\emph{Proc. 18th ACM SIGKDD
  Int. Conf. Knowl. Discovery {\&} Data Mining (KDD'12)}}.
  \bibinfo{address}{Beijing, China}, \bibinfo{pages}{696--704}.
\newblock


\bibitem[Zhou et~al\mbox{.}(2018)]%
        {zhou2018deeplink}
\bibfield{author}{\bibinfo{person}{Fan Zhou}, \bibinfo{person}{Lei Liu},
  \bibinfo{person}{Kunpeng Zhang}, \bibinfo{person}{Goce Trajcevski},
  \bibinfo{person}{Jin Wu}, {and} \bibinfo{person}{Ting Zhong}.}
  \bibinfo{year}{2018}\natexlab{}.
\newblock \showarticletitle{{DeepLink}: A deep learning approach for user
  identity linkage}. In \bibinfo{booktitle}{\emph{Proc. 37th IEEE Conf. Comput.
  Commun. (INFOCOM'18)}}. \bibinfo{address}{Honolulu, HI},
  \bibinfo{pages}{1313--1321}.
\newblock


\end{thebibliography}










\end{document}